\documentclass[12pt,preprint]{aastex}
\usepackage{graphicx}
\usepackage{natbib}

\shorttitle{Metal-Poor Planet Search. II.}
\shortauthors{Sozzetti et al.}

\begin{document}


\title{A Keck HIRES Doppler Search for Planets Orbiting Metal-Poor Dwarfs. 
II. On the Frequency of Giant Planets in the Metal-Poor Regime}


\author{Alessandro Sozzetti\altaffilmark{1,2}, 
 Guillermo Torres\altaffilmark{1}, David W. Latham\altaffilmark{1},
Robert P. Stefanik\altaffilmark{1}, Sylvain G. Korzennik\altaffilmark{1}, 
Alan P. Boss\altaffilmark{3}, Bruce W. Carney\altaffilmark{4}, and John
B. Laird\altaffilmark{5}} 
\altaffiltext{1}{Harvard-Smithsonian Center for Astrophysics, 60
Garden Street, Cambridge, MA 02138 USA}
\altaffiltext{2}{INAF - Osservatorio Astronomico di Torino, 10025 Pino 
Torinese, Italy}
\altaffiltext{3}{Department of Terrestrial Magnetism, 
Carnegie Institution of Washington, 
5241 Broad Branch Road, NW, Washington, DC 20015 USA}
\altaffiltext{4}{Department of Physics \& Astronomy,
University of North Carolina at Chapel Hill, Chapel Hill, NC 27599 USA}
\altaffiltext{5}{Department of Physics \& Astronomy,
Bowling Green State University, Bowling Green, OH 43403 USA}
\email{asozzett@cfa.harvard.edu}
\email{gtorres@cfa.harvard.edu}
\email{dlatham@cfa.harvard.edu}
\email{skorzennik@cfa.harvard.edu}
\email{rstefanik@cfa.harvard.edu}
\email{boss@dtm.ciw.edu}
\email{bruce@physics.unc.edu}
\email{laird@bgsu.edu}


\begin{abstract}

We present an analysis of three years of precision radial velocity measurements 
of 160 metal-poor stars observed with HIRES on the Keck 1 telescope. 
We report on variability and long-term velocity trends for each star in our 
sample. We identify several long-term, low-amplitude radial-velocity variables 
worthy of follow-up with direct imaging techniques. 
We place lower limits on the detectable companion mass as a function of orbital 
period. Our survey would have detected, with a $99.5\%$ confidence level, over 
$95\%$ of all companions on low-eccentricity orbits with velocity semi-amplitude 
$K \gtrsim 100$ m s$^{-1}$, or $M_p\sin i\gtrsim 3.0\,M_\mathrm{J}(P/\mathrm{yr})^{(1/3)}$, 
for orbital periods $P\lesssim 3$ yr. 
None of the stars in our sample exhibits radial-velocity variations compatible 
with the presence of Jovian planets with periods shorter than the survey duration. 
The resulting average frequency of gas giants orbiting metal-poor dwarfs with 
$-2.0\lesssim$[Fe/H]$\lesssim -0.6$ is $f_p<0.67\%$ (at the $1\sigma$ confidence level). 
We examine the implications of this null result in the context of the observed 
correlation between the rate of occurrence of giant planets and the metallicity 
of their main-sequence solar-type stellar hosts. By combining our dataset with the 
Fischer \& Valenti (2005) 
uniform sample, we confirm that the likelihood of a star to harbor a planet more massive 
than Jupiter within 2 AU is a steeply rising function of the host's metallicity. 
However, the data for stars with $-1.0\lesssim$[Fe/H]$\lesssim 0.0$ are compatible, 
in a statistical sense, with a constant occurrence rate $f_p\simeq 1\%$. Our results 
can usefully inform theoretical studies of the process of giant planet formation across 
two orders of magnitude in metallicity.

\end{abstract}





\section{Introduction}

Fourteen years after the Doppler detection of the Jupiter-mass planet
around the nearby, solar-type star 51 Peg (Mayor \& Queloz 1995), 
extrasolar planet discoveries have been reported by teams using four 
different techniques. Decade-long, high-precision (3-5 m s$^{-1}$) radial-velocity 
surveys of $\sim 2500$ F-G-K dwarfs and sub-giants in the solar neighborhood
\footnote{For reviews of the status of the most successful Doppler surveys 
for planets see for example Butler et al. (2006) and Udry \& Santos (2007) 
and references therein, or Jean Schneider's website {\tt http://exoplanet.eu}~.} 
($d\lesssim 30-50$ pc) have yielded so far the overwhelming majority 
of the objects in the present sample (340 planets in $\sim270$ systems, 
as of February 2009). Ground-based photometric transit surveys (for a review see 
for example Charbonneau et al. 2007, and references therein) are now 
uncovering new transiting systems at a rate of $\sim 20$ per year, and there 
are very bright prospects for an even steeper increase in the detection rate 
thanks to large surveys with space-borne observatories, which are currently 
ongoing ($CoRoT$; Baglin et al. 2002) or will start in the near future ($Kepler$; Borucki et al. 2003). 
Finally, an additional dozen or so sub-stellar companions have also 
been detected recently by means of gravitational microlensing 
(e.g., Bond et al. 2004; Beaulieu et al. 2006; Gaudi et al. 2008) 
as well direct imaging (e.g., Chauvin et al. 2005; Neuh\"auser et al. 2005; Kalas et al. 2008; 
Marois et al. 2008) surveys, or timing of stellar oscillations (Silvotti et al. 2007; 
Lee et al. 2009). 

The aims of Doppler surveys for planets have evolved in the recent past. 
On the one hand, existing surveys are extending their time baseline and/or 
are achieving higher velocity precision ($\lesssim 1$ m s$^{-1}$, see for 
example Lovis et al. 2006), to continue searching for planets at increasingly 
larger orbital distances (e.g., Fischer et al. 2007) and with increasingly smaller masses 
(e.g., Udry et al. 2007; Mayor et al. 2008). 
On the other hand, early evidence for a strong relationship between the 
physical properties of stars and the likelihood that they harbor planets 
has prompted both theoretical analyses attempting to reconcile the 
observed trends within the framework of planet formation models as well 
as renewed experimental efforts to put such trends on firmer statistical 
grounds and thus thoroughly test the theoretical explanations put 
forth to explain their existence. 

For example, if the surface density of solids in a protoplanetary 
disk is proportional to the mass of the central star, then, 
within the framework of the core accretion model of giant 
planet formation (e.g., Pollack et al. 1996; Alibert et al. 2005), one would 
expect a positive correlation between the mass of the host and 
the occurrence rate and mass of planets (Laughlin et al. 2004; Ida \& Lin 2005), 
so that low-mass stars should have a much lower frequency of Jupiter-mass 
companions, while low-mass planets (Neptune-like and 
terrestrial-type) around these stars should be relatively common. 
Note, however, that if the disk initial conditions are independent of 
stellar mass, then opposite conclusions are reached (Kornet et al. 2006). 
Observational evidence so far collected by dedicated Doppler surveys 
of a few hundred bright M dwarfs (Butler et al. 2004; Rivera et al. 2005; 
Bonfils et al. 2005; Endl et al. 2006) appears to support the former scenario.
Alternatively, both large-separation gas giants as well as ``super-Earths'' 
formed via the disk instability mechanism (e.g., Durisen et al. 2007, 
and references therein) might be found around low-mass stars (Boss 2006a,b).
In this scenario, the inner (hot) super-Earths found by Doppler surveys 
(Rivera et al. 2005; Udry et al. 2007; Mayor et al. 2008; Forveille et al. 2008) would still
have formed by collisional accumulation, while the outer (cold) super-Earths found
by microlensing surveys (e.g., Beaulieu et al. 2006; Bennett et al. 2008) 
would have formed by disk instability followed by photoevaporative stripping 
of their gaseous envelopes by EUV/FUV radiation from nearby massive stars (Boss 2006b).
The prediction of theoretical models that around massive stars planets 
might be more common, and more massive, is presently being investigated 
by dedicated surveys of Hertzsprung gap sub-giants (Johnson et al. 2006), 
heavily evolved stellar samples belonging to the red-giant branch 
and clump regions of the H-R diagram (Frink et al. 2002; 
Sato et al. 2003; Setiawan et al. 2003; Hatzes et al. 2005; Lovis \& Mayor 2007; 
Niedzielski et al. 2007, and references therein) and early-type dwarfs (Galland et al. 2005).

The other important relationship uncovered so far 
between planet characteristics and frequencies and host properties is 
quantified by the strong dependence of planet occurrence rates on stellar 
metallicity. The observational evidence (Gonzalez 1997; Santos et al. 2004; 
Fischer \& Valenti 2005) has found theoretical support within the 
context of the core accretion model (Ida \& Lin 2004; Kornet et al. 2005). 
As increased surface density of solids would facilitate the growth 
of embryonic cores, and as stellar metallicity is naturally thought of as 
a proxy for the actual heavy metal content of the primordial circumstellar 
disk, then gas giant planet formation is greatly enhanced around more 
metal-rich stars. The alternative view of gas giant formation by 
local gravitational instability, instead, is less sensitive to the 
disk metal content, and as a consequence giant planets formed via this 
mechanism should be found orbiting metal-rich and 
metal-poor stars with roughly equal probability (Boss 2002). 
However, Mayer et al. (2007) have found that higher metallicity may
encourage giant planet formation by disk instability because an
increased mean molecular weight leads to lower gas pressure.
The observed trend for [Fe/H]$\gtrsim 0.0$ is usually taken as strong
support for the core accretion model, though recently evidence has
emerged from surveys of giant stars (Pasquini et al. 2007) that
the metallicity correlation is due at least in part to pollution
of the surface layers of dwarf stars by ingested planetary materials.
Doppler surveys biased toward high-metallicity stellar samples 
(Fischer et al. 2005; Bouchy et al. 2005) have begun in recent years, 
prompted by the enhanced chances of finding large numbers of planets. 
In order to better characterize the dependence of giant planet frequency 
in the metal-poor regime, two surveys have started monitoring 
stellar samples with [Fe/H]$\lesssim -0.5$. A southern sample of 
$\sim 100$ stars is being observed by the Geneva team (Mayor et al. 2003, 
http://www.eso.org/observing/proposals/gto79/harps/4.txt), while 
our group (Sozzetti et al. 2006, S06 hereafter) has focused on a 
northern sample ($\delta\gtrsim -25^o$) of $\sim 200$ objects. 

In S06 we described our dedicated Doppler survey of metal-poor dwarfs 
with HIRES on Keck 1, and presented an assessment of the overall quality 
of our measurements, making use of the results for stars that showed no 
velocity variations. Here we report on results for all stars observed 
with Keck/HIRES to date. 
The structure of our second paper is the following. Sect. 2 provides 
a brief overview of the Keck/HIRES Doppler survey for giant planets 
orbiting low-metallicity main-sequence stars. We present in \S~3 results for all 
the stars observed in our program. Sect. 4 is dedicated to a statistical 
analysis of all the data, to assess variability thresholds and the presence 
of linear and non-linear trends on a star-by-star basis. In \S~5 we 
place quantitative limits on the detectable companion mass as a function 
of orbital period. Finally, in \S~6 we discuss our results, and their 
implications for the frequency of gas giant planets in the metal-poor 
regime and for the observed correlation between stellar metallicity and 
planet occurrence; we compare our findings to the predictions of proposed 
models of giant planet formation, and we suggest venues to further our 
overall understanding of this fundamental issue of planetary system science.

\section{The Keck/HIRES Doppler survey of metal-poor dwarfs}

\subsection{Target sample}

Table~\ref{taball} lists 160 metal-deficient stars observed at least twice 
during our three-year long observing campaign. The list of targets includes spectral type, 
visual magnitude $V$, effective temperature $T_\mathrm{eff}$, metallicity [Fe/H], 
stellar mass $M_\star$, $B-V$ and $V-K$ colors, surface gravity $\log g$, and 
distance estimate $d$ (based on their $Hipparcos$ parallax when available, photometric otherwise). 
The complete list included 200 stars, but for 20\% of the sample the non-optimal 
timing of the observing nights at Keck did not allow us to collect multiple exposures before the 
program was ended. About 10\% of the targets in Table~\ref{taball} are also in the list of 
sourthern metal-poor stars observed with HARPS, and a similar fraction has also been observed 
by the California team (with about 50\% of these subsamples appearing in the target lists of all three 
groups).

\begin{figure}
\centering
\includegraphics[width=.65\textwidth]{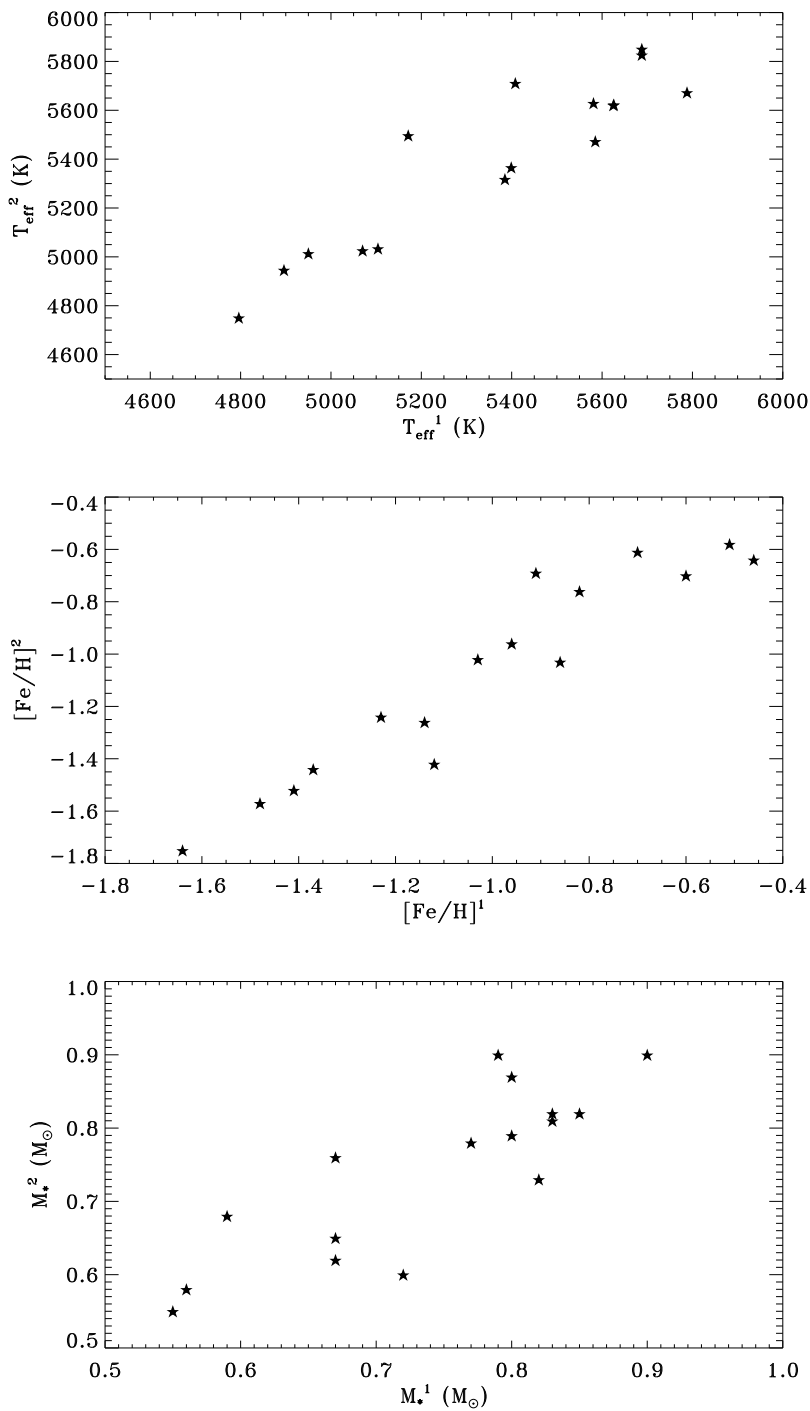}
\caption{Comparison between the values of $T_\mathrm{eff}$, [Fe/H], and $M_\star$ 
(top, center, and lower panel, respectively) for the objects in common between our program stars 
(with superscript 2 on the $y$-axis) and the SPOCS database (with superscript 1 on the $x$-axis) 
described in Valenti \& Fischer (2005) and Takeda et al. (2007). 
\label{compatmpar}}
\end{figure}

As described in S06, all stars are drawn from the Carney-Latham and Ryan samples of metal-poor, 
high-velocity field stars (e.g., Carney et al. 1994; Ryan 1989; Ryan \& Norris 1991). 
Based on a decade-long radial-velocity monitoring with the CfA Digital Speedometers (Latham 1992) 
none of our program stars showed signs of velocity variation at the 0.5 to 1.0 km s$^{-1}$ level. 
Additional target selection criteria are driven by the need to find a balance between the 
sample size, the amount of observing time devoted to each target, 
and the achievability of a radial-velocity (RV) precision good enough 
for planet detection, and this ultimately imposed constraints on the properties of the 
target sample. In the end, based on detailed simulations with the CfA library of synthetic 
stellar spectra (see S06 for details), we selected targets with $V\lesssim 12$ mag, 
$T_\mathrm{eff}\lesssim 6000$ K, and $-2.0\lesssim$[Fe/H]$\lesssim -0.6$. Table~\ref{taball} 
lists photometric temperatures and spectroscopic metallicities determined from 
comparison of observed and synthetic spectra, and stellar masses derived from 
comparison with the Yonsei-Yale stellar evolution models (Demarque et al. 2004) in a 
color-magnitude diagram (the best agreement over a wide range of [Fe/H] values was obtained 
using the $V-K$ colors). The analysis closely follows the procedures of Carney et al. (1987, 1994). 
Notable changes with respect to that approach 
include the determination of photometric $T_\mathrm{eff}$ values using the 
Alonso et al. (1996) calibrations and the adoption of reddening values based on 
$uvby\beta$ photometry whenever possible instead of the default approach of using reddening maps 
from Burstein \& Heiles (1982). All new details of the procedure as applied to all stars included 
in the proper motion survey of Carney et al. (1994) will be discussed extensively 
in a forthcoming paper (Carney et al., in preparation). 
As far as the subset of metal-poor stars presented in Table~\ref{taball} is concerned, 
typical uncertainties on $T_\mathrm{eff}$, [Fe/H], and $M_\star$ are 100 K, 0.1 dex, and 
0.1 $M_\odot$ respectively. In order to verify the reliability of our internal errors on the 
stellar parameters we have carried out an independent check comparing our temperature, metallicity, 
and mass estimates with those reported in the SPOCS catalog (Valenti \& Fischer 2005; Takeda et al. 2007) 
for the subset of 16 stars in common between our program and the one of the California team. 
The result of the comparison is shown in Figure~\ref{compatmpar}. The agreement between the two 
sets of data is broad: A straight line fit in three cases gives slopes of 0.99, 1.05, and 0.89 
for the top, center, and lower panel, respectively. The mean differences in the 
three parameters are $-37\pm33$ K, $0.04\pm0.03$ dex, and $-0.002\pm0.016$ $M_\odot$, 
respectively, with standard deviations of $134$ K, $0.12$ dex, and $0.06$ $M_\odot$, respectively. 
Although based on a relatively small subset of common stars, these findings indicate that our values 
are compatible with those reported in the SPOCS catalog within our quoted errors, 
reinforcing our confidence on the reliability of the procedures we adopted to determine the 
stellar parameters and related uncertainties for our program stars. Updated estimates of the stellar parameters 
and detailed spectroscopic abundance determinations for this sample 
(e.g., including [Fe/H] and [$\alpha$/Fe]), based on our high-resolution, high-$S/N$ Keck spectra, 
will be the main focus of future work.

\subsection{Observations}

All observations were collected with HIRES and its I$_2$ cell (except for one Iodine-free 
template exposure per target) on the Keck 1 telescope (Vogt et al. 1994). 
The spectrometer slit used at Keck was $0.57\arcsec$. During the observing runs 
in 2003 and 2004, with the old HIRES CCD setup the nominal resolving power and spectral coverage 
were $\sim 65,000$ and 3850-6200 \AA, respectively. These figures, after the summer 2004 upgrade 
of the HIRES CCD, became $\sim 71,000$ and 3200-8800 \AA, respectively. As described in S06, 
exposure times were calculated on a star by star basis, with the aim of reaching uniform RV precision 
($\approx 10$ m s$^{-1}$) for the whole sample. The resulting average exposure time, yielding a 
typical $S/N$ ratio of $\sim 100$ pixel$^{-1}$, is ~9 min. 
Integration times were limited to a maximum of 
15 minutes, in order to minimize timing uncertainties and subsequent systematic errors 
in the barycentric velocity correction. 

\subsection{Analysis}

To extract the RV information from each star+iodine spectrum, we perform a full spectral
modeling which includes the reconstruction of the asymmetries, spatial and temporal variations 
in the HIRES instrumental profile at the time of observation. First, the spectrum is subdivided 
into smaller segments, and then for each segment a multi-parameter, $\chi^2$-minimization 
scheme is performed and a best-fit model is obtained, based on a 
template (iodine-free) spectrum of the target and a laboratory spectrum of the I$_2$ cell 
(e.g., Valenti et al. 1995; Butler et al. 1996). 
The RV of the target is measured with respect to the template, while the superimposed I$_2$ lines 
provide the wavelength calibration information. Uncertainties in the RV measurements are determined 
from the scatter about the mean for each spectral order divided by the square root of the 
number of orders containing I$_2$ lines. Our algorithm for precision RV measurements 
follows a procedure based on the methodology developed for the AFOE spectrograph (Korzennik et al. 2000),
and adapted for the processing of HIRES spectra, as described in Sozzetti et al. (2008). 

\section{Radial-velocity results}

We summarize in Table~\ref{totrv} our radial-velocity results by giving, 
for each star, the number of observations, 
the total rms RV scatter, the mean internal uncertainty, and the duration of monitoring. 
All radial velocities not presented in this work are available upon request from A. Sozzetti.

As discussed in S06, the average RV scatter of the whole sample 
is $\bar\sigma_\mathrm{rms}\simeq 9$ m s$^{-1}$. No statistically significant 
trends of $\sigma_\mathrm{rms}$ with $V$, $T_\mathrm{eff}$, or [Fe/H] are 
present in the data, an indication that our predicted exposure times 
based on simulations with the CfA library of synthetic stellar spectra 
were reliable.  
Inspection of Table~\ref{totrv} shows that, while most of the objects in the 
sample have RV scatter comparable to the internal error estimates, 
several stars exhibit $\sigma_\mathrm{rms}$ larger (in some cases, much larger) 
than $\sigma_\mathrm{int}$. It is then necessary to perform a thorough statistical 
examination of the complete RV dataset, testing for variability and long-term 
trends (given the limited and typically very sparse number of spectra per star, 
a search for significant periodic signals is not feasible).

\subsection{Statistical Analysis: Testing for excess variability}

Almost all stars in our sample belong to the halo or thick-disk populations 
(Carney et al. 1994, 1996). In particular, population membership probabilities calculated 
following the methods described in Bensby et al. (2003), 
Venn et al. (2004), and Reddy et al. (2006) indicate that 87\% of the stars observed 
in the Keck/HIRES survey (138 out of 160) have less than 20\% chances of belonging 
to the thin disk of the Milky Way. The fact that these stars are likely members of 
the thick-disk and halo populations also implies that they typically have very old ages 
(e.g., Holmberg et al. 2007). An advantage of dealing with rather old main-sequence 
stars is that they tend to be slow rotators (all our program stars have 
projected rotational velocities $v\sin i\leq 10$ km s$^{-1}$), 
and exhibit low levels of chromospheric emission. Both rotation and stellar 
activity constitute sources of intrinsic radial
velocity ''jitter'' that can mask, and sometimes even mimic, the
presence of orbital reflex motion due to planetary mass companions
(Saar et al. 1998; Santos et al. 2000; Queloz et al. 2001;
Paulson et al. 2004), particularly in the case of young (e.g., Setiawan et al. 2008; Hu\'elamo et al. 2008) 
and evolved (e.g., Carney et al. 2003, 2008a, 2008b; Hekker et al. 2008) stars. 
For the targets observed in our survey, sources of intrinsic RV variability should then 
be minimal. For this reason, we did not 
add in quadrature to the internal error estimates any terms representing intrinsic 
stellar variability. We then ask if the scatter $\sigma_\mathrm{rms}$ in the velocities is consistent 
with our $\sigma_\mathrm{int}$ estimate for each star. We perform three tests. 

First, we perform an $F$-test, comparing the ratio of variances 
$\sigma^2_\mathrm{rms}/\sigma^2_\mathrm{int}$. Next, we use the distribution 
of the reduced $\chi^2_\nu$ (with $\nu=N-m$ being the number of degrees 
of freedom, and $m=1$ for a constant model) to test if the velocities are 
consistent with being drawn from a normal distribution with variance 
$\sigma^2_\mathrm{int}$. Finally, we determine whether the underlying one-dimensional 
probability distributions of $\sigma_\mathrm{rms}$ and $\sigma_\mathrm{int}$ 
are significantly different, based on Kuiper's $Ku$ test. This test is preferable 
to the more widely known Kolmogorov-Smirnov ($K-S$) test, as while both tests 
determine the maximum deviation between two cumulative distribution functions, 
the former is equally sensitive in the tails as at the median. This test is carried out only 
for objects with $N > 3$, as for very small datasets it is not meaningful. In all cases, 
excess variability in the dataset for any given star can be assessed, in a statistical 
sense, when a small value of the false alarm probability (Pr($F$), Pr($\chi^2_\nu$), 
and Pr($Ku$), respectively) indicates that the null hypothesis should be rejected 
(at a given confidence level). Given the relatively small number of observations per 
target, to account for the possibility of outliers in the data, and to minimize the 
likelihood of obtaining more than 1 false signal in our sample, we have decided 
to adopt high thresholds of the false-alarm probabilities ($Pr < 0.005$, i.e. a 
99.5\% confidence level) and to request stringent agreement between the three tests, 
before any given object can be flagged as very likely variable. 

We report in Table~\ref{stattests} the results of the variability tests 
applied to the entire sample. Nine stars, with as few as 3 and as many as 
11 observations, are identified as variables based on the above criteria. 
These nine objects (HD 7424, G197-45, G 237-84, G 63-5, G 135-46, HD 192718, HD 210295, 
G 27-44, and G 28-43) all exhibit $\sigma_\mathrm{rms}/\bar\sigma_\mathrm{int} \gtrsim 4$, 
i.e. a scatter in the measurements at least four times larger than the nominal average internal error. 
Ten stars are recognized as significantly variable by two out of three tests, and these comprise 
objects with $2.0\lesssim\sigma_\mathrm{rms}/\bar\sigma_\mathrm{int}\lesssim 3.0$. 
Twenty two stars fail only one test, and for the remaining 119 objects all tests return 
$Pr > 0.005$. All radial velocities for stars not deemed significantly variable, i.e. 
corresponding to objects with $\sigma_\mathrm{rms}/\bar\sigma_\mathrm{int}\lesssim 3.0$, 
are presented in Table~\ref{rvsumm}.

\subsection{Long-term linear and curved trends: Individual objects}

We next investigate the possibility that the RV data for each object
can be better described by a linear slope, which 
would indicate the presence of a massive, distant companion orbiting with a period 
greatly exceeding the duration of the observations ($>> 3$ yr). Such long-term trends 
could in fact directly affect our search for planetary-mass companions with $P < 3$ yr, 
and thus should be immediately subtracted if present.  

For each star with at least three RV observations, 
we fit a straight line to the measured velocities. We derive a best-fit 
slope and its uncertainty, then verify its significance using an $F$-test that compares 
the weighted sum of the squares of the residuals from the straight line fit to that about 
the mean. Large values of $F$ (and correspondingly low values of $P(F)$) would indicate 
that the data are better described by a straight line, i.e. they display a significant 
long-term linear trend. For the purpose of this study, we use again the prescription 
($Pr(F) < 0.005$) to discriminate between statistically significant and insignificant 
slopes. We conclude by re-computing $\sigma_\mathrm{rms}/\bar\sigma_\mathrm{int}$ and $Pr(\chi^2_\nu)$ 
on the post-fit residuals, to determine whether the two-parameter model can or cannot be 
considered satisfactory (even in cases when a significant slope is found). The results are 
summarized in Table~\ref{trendstest}. 

\begin{figure}
\plotone{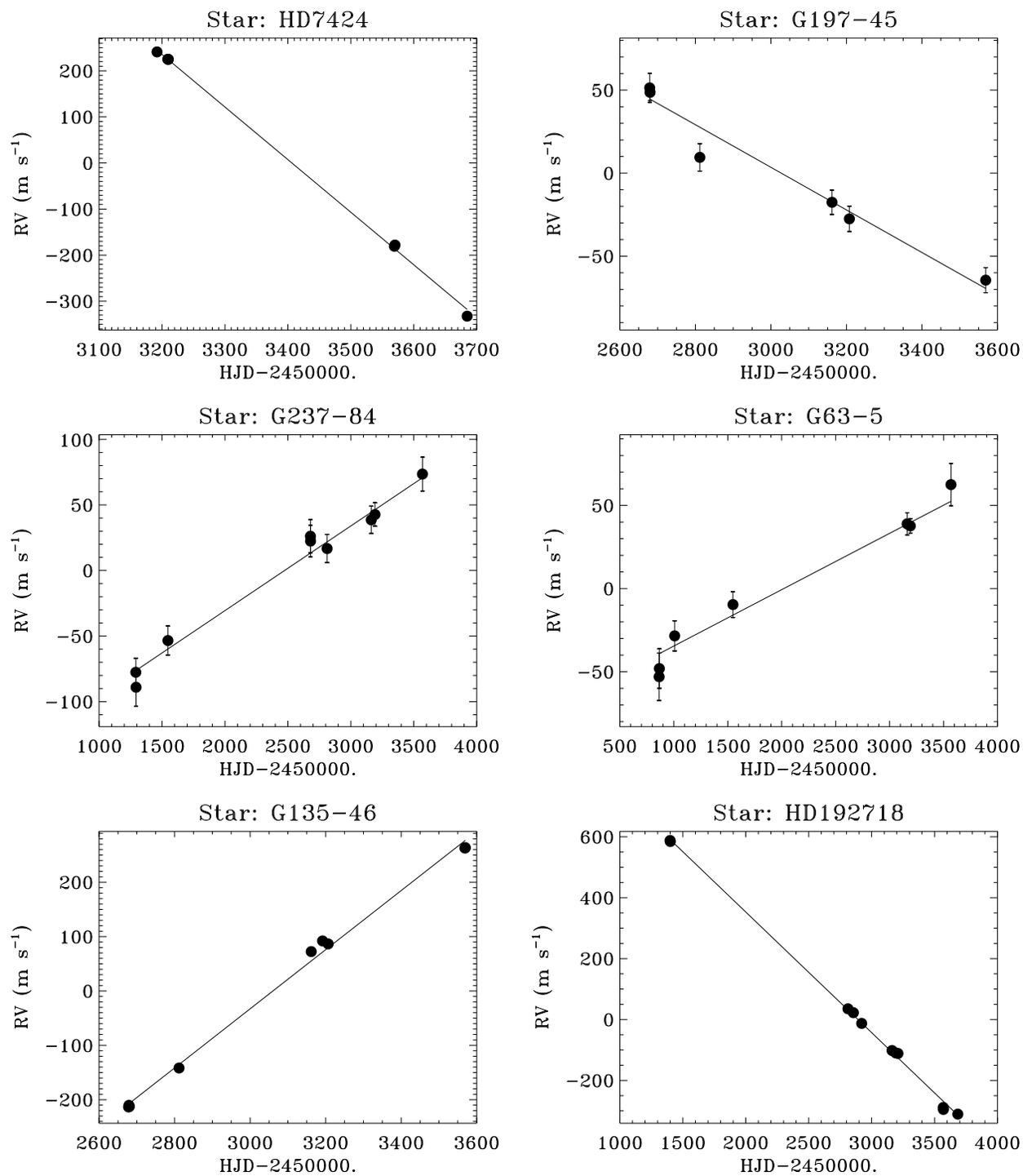}
\caption{Linear fits (solid lines) to the RV data of six RV variables: 
HD 7424, G 197-45, G 237-84, G 63-5, G 135-46, and HD 192718.\label{lintrend1}}
\end{figure}

\begin{figure}
\centering
\includegraphics[width=.75\textwidth]{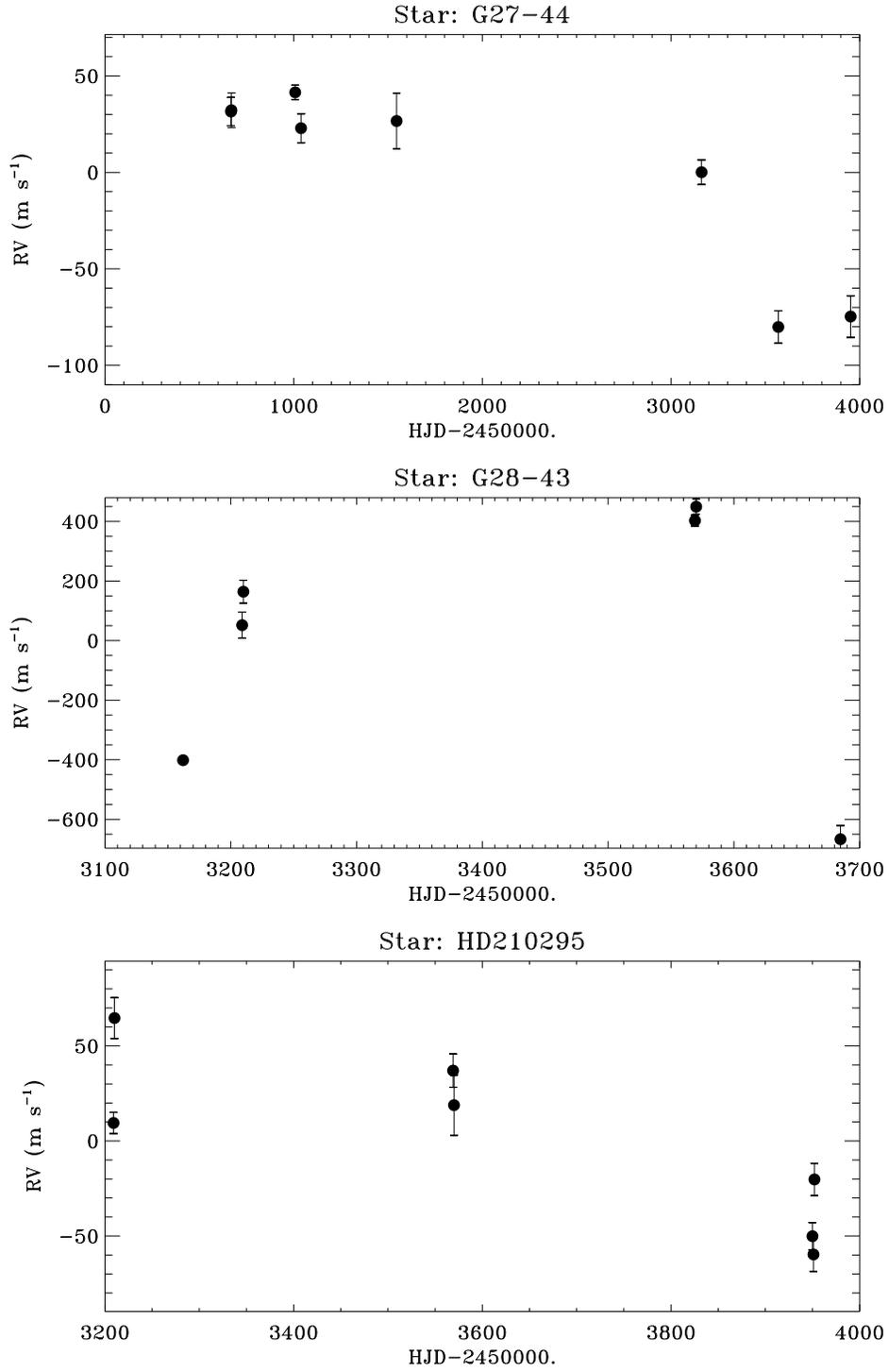}
\caption{Radial velocity as a function of time for three stars exhibiting significant 
non-linear RV variability: G 27-44 (upper panel), G 28-43 (middle panel), and 
HD 210295 (lower panel).\label{lintrend2}}
\end{figure}

For most stars, insignificant slopes are derived. For six of the nine stars that were flagged 
as clearly variable in the previous section $(\sigma_\mathrm{rms}/\bar\sigma_\mathrm{int} \gtrsim 4$), 
the slope is quite significant, and after subtraction of the linear RV trends the post-fit residuals agree 
much more closely with the measurement errors. We show the best-fit linear trends for these objects (HD7424, 
G197-45, G237-84, G63-5, G135-46, HD192718) in Figure~\ref{lintrend1}. 
We discuss the results for these stars collectively below. 
As for the other three variables (HD210295, G27-44, and G28-43), the RV residuals are not 
significantly improved after fitting for a linear trend. We show in Figure~\ref{lintrend2} 
the RVs collected for the three stars. They all clearly exhibit significantly non-linear RV variations. 
The results for each of these stars are discussed separately below. 

None of the 32 stars which fail either one or two of the three statistical tests for variability appears to 
exhibit significant ($Pr(F) < 0.005$) long-term trends. However, a significant slope is found for 
one of the stars (G204-30) which do not appear as RV variables based on all tests. The 
results for this object are also discussed below. 

\subsubsection{Stars with secular trends}

The seven stars with significant long-term linear slopes 
(HD 7424, G 197-45, G 237-84, G 63-5, G 135-46, HD 192718, and G 204-30) 
are all unevolved dwarfs with $-1.0\lesssim$[Fe/H]$\lesssim -0.6$ (see 
Table~\ref{taball}). In the five cases that are bright enough to have been
observed by Hipparcos, the single-star astrometric solutions show no
indication of excess scatter, suggesting the RV trends we see are
likely due to a brown-dwarf or stellar companion in a very long-period
orbit. One of these, G 63-5, is listed in the Washington Double Star Catalog 
as having a common proper motion M0 binary companion at $\sim 80\arcsec$, 
but this is likely to be too distant to be the cause of the observed slope 
(at a distance $d\sim60$ pc, this corresponds to a linear orbital semi-major axis of $~\sim5000$ AU). 
All radial velocities collected for these objects are presented in Table~\ref{rvlin}.

\subsubsection{G 27-44}

This relatively metal-poor ([Fe/H]$\simeq -0.80$) 
dwarf was observed by $Hipparcos$, which did not detect any significant 
deviations from a single-star astrometric model. 
The object was included in the $I-$band CCD imaging search for wide metal-poor 
binaries of Zapatero Osorio \& Mart{\`\i}n (2004), who searched for companions 
within $\sim 25\arcsec$ of 473 metal-deficient stars. This survey was sensitive to 
nearby companions up to 5 mag fainter than the target star. No indication of 
companions within the survey limits was found. We conclude this 
star is orbited by a previously undetected brown dwarf or stellar companion. 
The RV data collected for G 27-44 are presented in Table~\ref{rvnolin}. 
Direct imaging observations in the infrared are ongoing to ascertain its nature. 

\subsubsection{G 28-43}

This low-metallicity ([Fe/H]$\simeq -1.80$) dwarf was observed by $Hipparcos$, but it is an apparent case of 
misidentification. The entry HIP 114349 includes the components 
CCDM J23096+0043 A and CCDM J23096+0043 B. The secondary (CCDM J23096+0043 B, $V=11.5$) 
is supposedly placed at $12.2\arcsec$ from the primary, at a position angle of 195 deg. 
However, Zapatero Osorio \& Mart{\`\i}n (2004) do not confirm it using their data 
and images from the Digitized Sky Surveys. On the other hand, Woolf \& Wallerstein 
(2005) argue that G 28-43 seems to be a double-lined spectroscopic binary. 
The presence in the spectrum of light from a close-in massive companion 
(with a separation $< 1\arcsec-2\arcsec$ would in fact explain the large error bars we 
obtain for the radial velocities (presented in Table~\ref{rvnolin}). 
This star thus appears to be orbited by a previously undetected 
brown dwarf or stellar companion. 
Direct imaging observations in the infrared are ongoing to ascertain its nature. 

\subsubsection{HD 210295}

This low-metallicity star ([Fe/H]$\simeq -1.50$) star was observed by $Hipparcos$, 
which did not detect any significant 
deviations from a single-star model. Again, Zapatero Osorio \& Mart{\`\i}n (2004) 
do not find evidence for companions up to 5 mag fainter within $\sim 25\arcsec$. 
The RV data collected for HD 210295 are presented in Table~\ref{rvnolin}. 
On the one hand, they appear compatible with the existence of a short-period companion inducing 
a relatively low velocity semi-amplitude ($\approx 50-60$ m s$^{-1}$). On the other 
hand, HD 210295 is the most evolved among the stars included in our survey. 
Hekker et al. (2008) have recently presented results from a multi-year precision 
RV monitoring campaign of $\sim 180$ K giants. They find a very significant 
inverse correlation between surface gravity and random RV `jitter', 
as well as a large number of stars with correlated, periodic RV variations that do not 
correlate with surface gravity. 
In particular, stars with $\log g\approx 2.8-3.0$, similar to HD 210295, 
often exhibit random half peak-to-peak RV variations of about $\sim 50-60$ m s$^{-1}$. 
Based on this evidence, it appears likely that the RV variations for HD 210295 
are of intrinsic nature, but we strongly encourage additional RV observations to 
better understand whether the RV jitter is clearly random or if it can be 
modeled with a periodic component.

\subsubsection{Summary}

Overall, our results indicate that none of the stars with clearly detected RV variations 
appear to be orbited by massive giant planet companions with periods shorter than the survey 
duration. Most of the variable stars exhibit long-term trends compatible with the existence 
of very distant brown dwarf or stellar companions.  
The only clear candidate exhibiting short-term, low-amplitude velocity variability is an 
evolved star, and the more likely explanation for the observed RV variability is to be 
found in some type of mechanism intrinsic to the star. Given the limited amount of data 
collected (typically, only a handful of observations per star over a time-span of three 
years), we can say very little about companions inducing RV variations close to the 
average internal errors we obtain ($\simeq 10$ m s$^{-1}$). 

\section{Limits on companion mass and period}

The quantitative determination of the sensitivity of an RV survey such as ours to 
planetary companions of given mass and period relies upon detailed numerical simulations 
of synthetic datasets of RV observations of a population of planetary systems of varying 
orbital properties and masses. The ability to recover, with a given level of statistical 
significance, the presence of a planetary signal or not translates into lower limits on the 
detectable (minimum) companion mass as a function of e.g. period and eccentricity (for an 
assumed mass of the central star). 

\subsection{Methodology}

Limits on companion detectability for a number of RV planet surveys have been presented in the past by, e.g., 
Walker et al. (1995), Cumming et al. (1999, 2008), and Endl et al. (2002, 2006). Nelson \& Angel (1998), 
Eisner \& Kulkarni (2001), Cumming (2004), and Narayan et al. (2005) have derived analytical expression for 
the detectability of a given RV signal in a variety of situations (large/small number of observations, 
periods shorter/longer than the duration of the observations, small and large companion masses). 
They used two different approaches, 
one based on a periodogram analysis to identify significant periodicities in a dataset, the other 
based on $\chi^2$- and $F$-tests to detect excess residuals above an assumed level of Gaussian noise. 

For our project, the periodogram approach is not feasible, due to the combined effect of very 
poor sampling and small number of observations per star. Thus, we rely on the alternate 
statistical approach. We start from the null hypothesis that there is no planet, i.e. that the 
RV of a given star is constant, and that the observed scatter is only due to measurement uncertainties. 
For each star in our program, we then add Keplerian signals to the data, at the exact times of observation, 
and perform a $\chi^2$-test with a 99.5\% confidence level (to minimize the number of 
false positives, given the sample size) on the new set of velocities, to verify if they are still consistent with being 
drawn from a Gaussian distribution with variance $\bar\sigma_\mathrm{int}$. Suppose $\chi^2_0$ is the 
value provided by the null model (no planet), then the test fails when $Pr(\chi^2\geq\chi^2_0)\leq 0.005$. 
In this case, the planet is considered detected. We employed the $\chi^2$-test to analyze $8\times 10^6$ 
star+planet systems, with the orbital elements and mass of the 
perturbing companion in the following ranges:  $0.01\leq P\leq 20$ yr, $0.0\leq e\leq 1.0$, 
$0.0\leq i\leq \pi/2$, $0.0\leq \omega\leq 2\pi$, $0.0\leq\tau\leq P$, 
$0.1\leq M_c\leq 20$ $M_J$. To transform any detected RV amplitude $K$ (in m s$^{-1}$) 
into a corresponding minimum companion mass (in $M_J$), we use the formula: 

\begin{equation}
M_c\sin i = \frac{K}{28.4\,\mathrm{m\, s^{-1}}}\left(\frac{M_\star}{M_\odot}\right)^{2/3}
            \left(\frac{P}{\mathrm{yr}}\right)^{1/3}\sqrt{1-e^2}
\end{equation}

In the simulations, the average stellar mass for our sample was utilized, 
i.e. $M_\star=0.80$ $M_\odot$. 

\subsection{Limits on companions and survey completeness}

\begin{figure}
\plotone{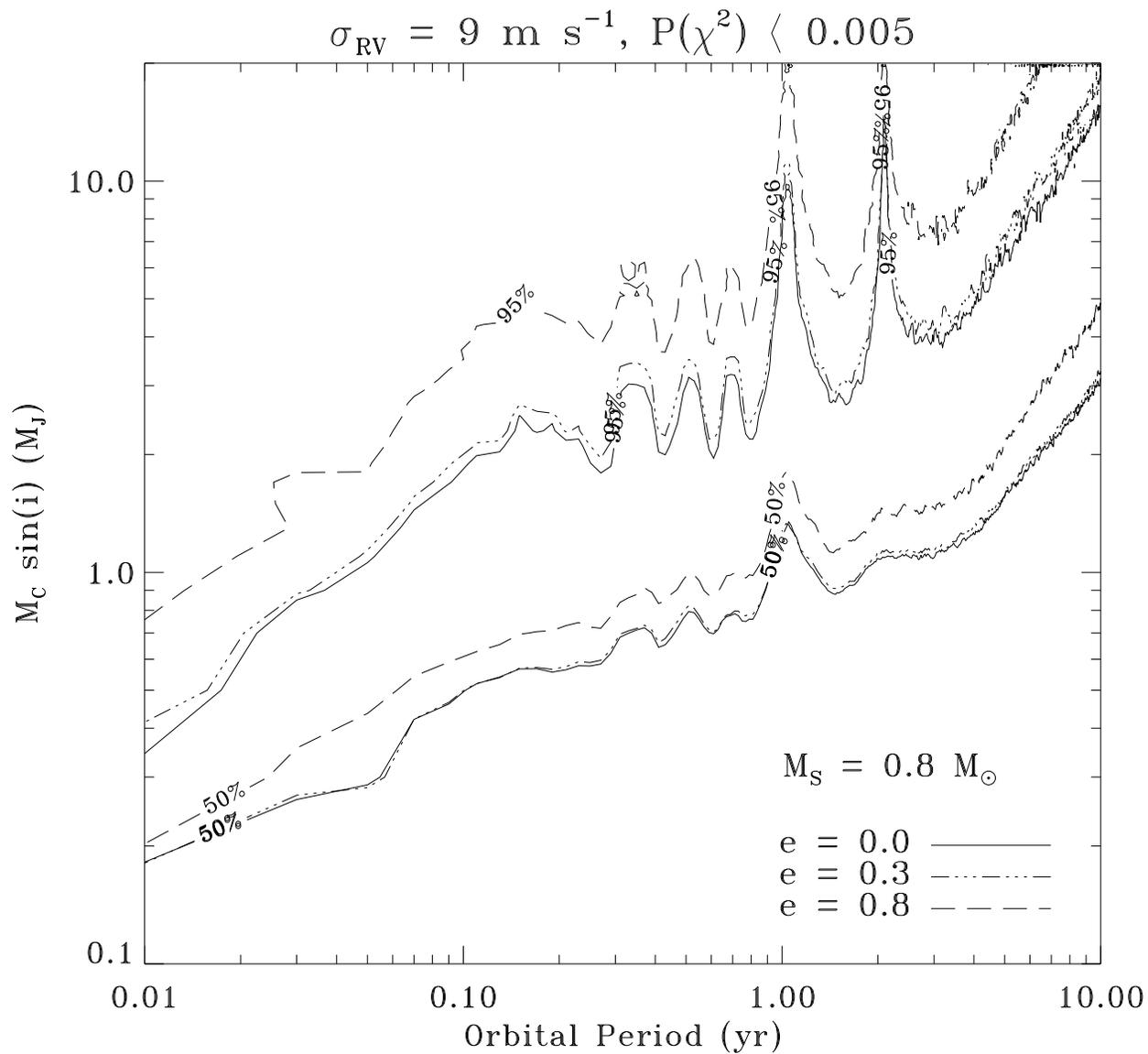}
\caption{Survey completeness for companions of given mass, orbital period, 
and eccentricity. The limits shown are for 50\% and 95\% completeness and for 
three realizations with different values of eccentricity.\label{detlim}}
\end{figure}

Figure~\ref{detlim} displays the survey sensitivity determined by this
method in the $M_p\sin i-P$ plane. The limits shown in the figure are for 50\% and 
95\% survey completeness, and for three realizations with different values of eccentricity: 
perfectly circular orbits, $e = 0.3$, corresponding to the median value of the $e$ 
distribution of known exoplanets, and $e=0.8$, which virtually encompasses the whole 
extrasolar planet sample. Note that the mass limits for the planets are somewhat 
conservative, because for less massive stars these companion mass values would be lower. 

The results shown in Figure~\ref{detlim} indicate that we would have detected 
virtually all companions on circular orbits with $M_c\sin i\gtrsim 3.0(P/1\,\mathrm{yr})^{1/3}$ $M_J$, 
and orbital periods $P \lesssim T$, with $T\approx 3$ yr being the average timespan 
of the observations. In this period range, the corresponding 
value of radial-velocity amplitude for secure detection is $K\approx 100$ m s$^{-1}$, 
or $K/\sigma_\mathrm{rms}\simeq 10$, if we use as a proxy for the average RV uncertainty 
of our data the median RV scatter of the whole sample $\sigma_\mathrm{rms}\simeq 9$ m s$^{-1}$. 
Instead, an amplitude such that $K/\sigma_\mathrm{rms}\simeq 3$ for $P < T$ (corresponding 
to $M_c\sin i\gtrsim 1.0(P/1\,\mathrm{yr})^{1/3}$ $M_J$), is recovered only 50\% of the time. 
Both results are in close agreement with those reported by e.g. Cumming (2004). 

Windows of poor detectability in period occur for two main reasons, i.e. due to the 
data structure (number of observations and phase coverage) and to the 1-yr seasonal period which 
affects any ground-based observing program. At these periods, the minimum detectable 
mass is anywhere between a factor of 1.5 (in the former case) and an order magnitude 
(in the latter case) larger. 

Orbital eccentricity has a very significant impact on the detection efficiency, as 
during the survey critical orbital phases of eccentric orbits can be missed due 
to sparse sampling, and a star with a very eccentric companion may show no appreciable 
RV variations. This effect is not prominent for mild eccentricities, up to values 
of $e$ close to the median of the exoplanet sample known to-date ($e = 0.3$). For 
very eccentric orbits (the case $e = 0.8$ is shown in Figure~\ref{detlim}), a 
much larger signal is required for secure detection. The equivalent sensitivity in 
mass degrades by a factor of a few. 

\begin{figure}
\plotone{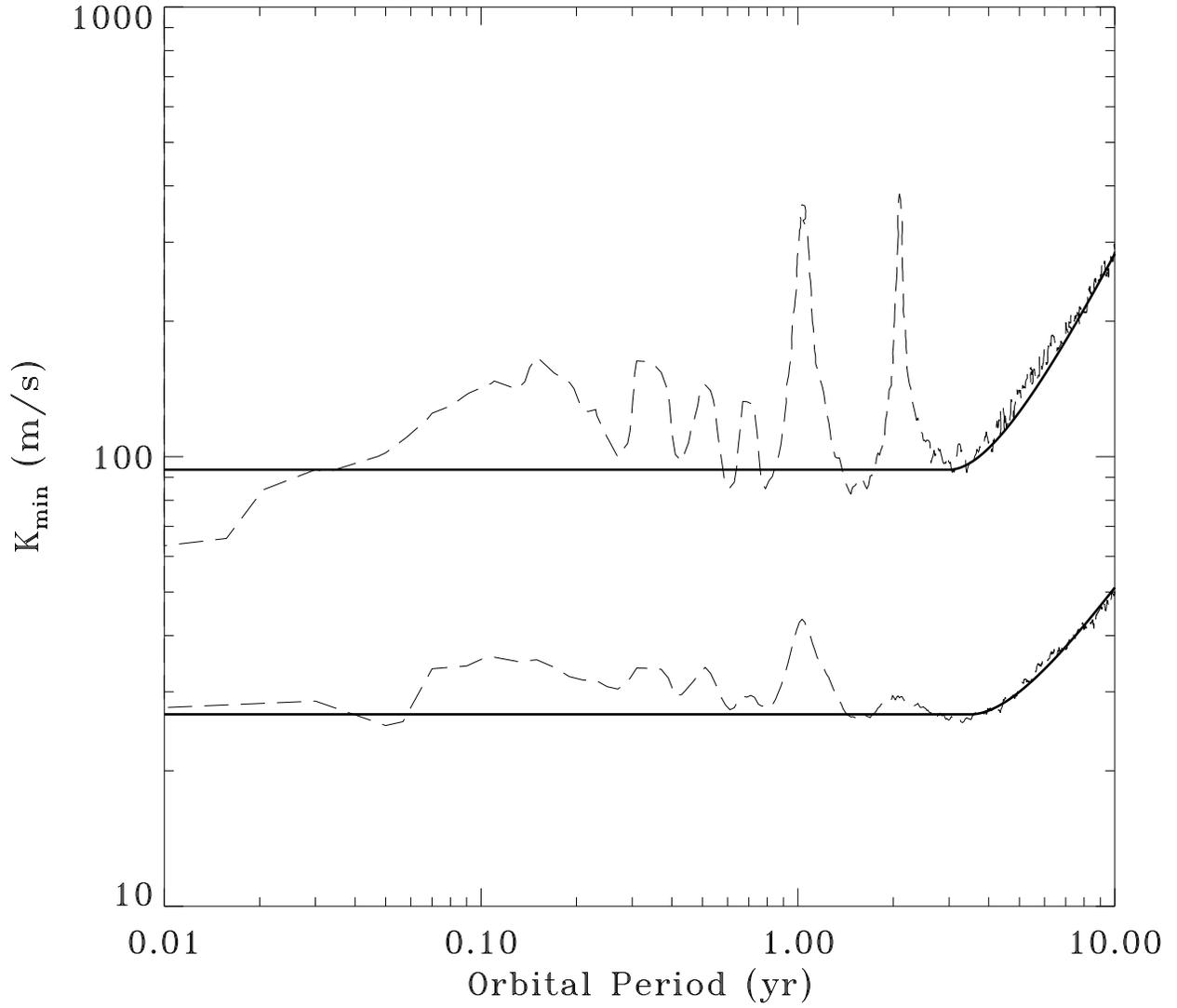}
\caption{Radial velocity detectability threshold for 50\% and 95\% 
detection efficiency (lower and upper dashed lines, respectively). The 
solid lines represent analytical results using the Cumming (2004) formalism.\label{rvlim}}
\end{figure}

For periods exceeding the time baseline of the observations ($P > T$), 
detection sensitivity drops quickly due to the fact that observations cover only a 
portion of an orbit. This is a well-known result, discussed by many groups in the 
past that addressed the issue of planet detection sensitivity both in the context of 
Doppler (see references at the beginning of this section) as well as astrometric 
(for a review, see Sozzetti 2005, and references therein) surveys for planets. 
For comparison, we show in Figure~\ref{rvlim}, as a function of $P$, 
the minimum RV amplitude $K_\mathrm{min}$ for 50\% and 95\% detection probability (dashed lines) 
obtained in the simulations superposed on the analytical results (solid thick lines) 
of Cumming (2004). In particular, the predicted behavior $K_\mathrm{min}\propto P$ for $P>T$ 
for the 50\% detection curve is very closely followed. For the case of 95\% 
completeness, we find a best-fit behavior $K_\mathrm{min}\propto P^{3/2}$, and 
for 99\% completeness (not shown) we correctly recover the steeper 
$K_\mathrm{min}\propto P^2$ behavior obtained by Cumming (2004). 

Based on the results of this and the previous section, we conclude that 
none of the metal-poor stars surveyed in our program reveal an excess RV scatter 
attributable to Keplerian reflex motion caused by a Jovian-mass companion with $P < T$.

\subsection{Frequency of close-in sub-stellar companions}

With this null result in hand, it is possible to draw conclusions on 
the frequency of giant planets orbiting metal-poor dwarfs to which our 
survey is sensitive (95\% completeness for $K > 100$ m s$^{-1}$, $P < 3$ yr, and $e < 0.3$). 
Given the relatively limited sample size, we elect to follow the 
approach described in e.g. Burgasser et al. (2003) and McCarthy \& 
Zuckermann (2004), rather than the more common procedure based on 
Poisson uncertainties. We then use the binomial probability 
distribution, which determines the probability $P(f_p;\,n,\,N)$ of $n$ detections 
given a sample of size $N$ when the true planetary companion frequency is $f_p$ as:

\begin{equation}
P(f_p;\,n,\,N) = \frac{N!}{n!(N-n)!}f_p^n(1-f_p)^{N-n}
\end{equation}

\begin{figure}
\plotone{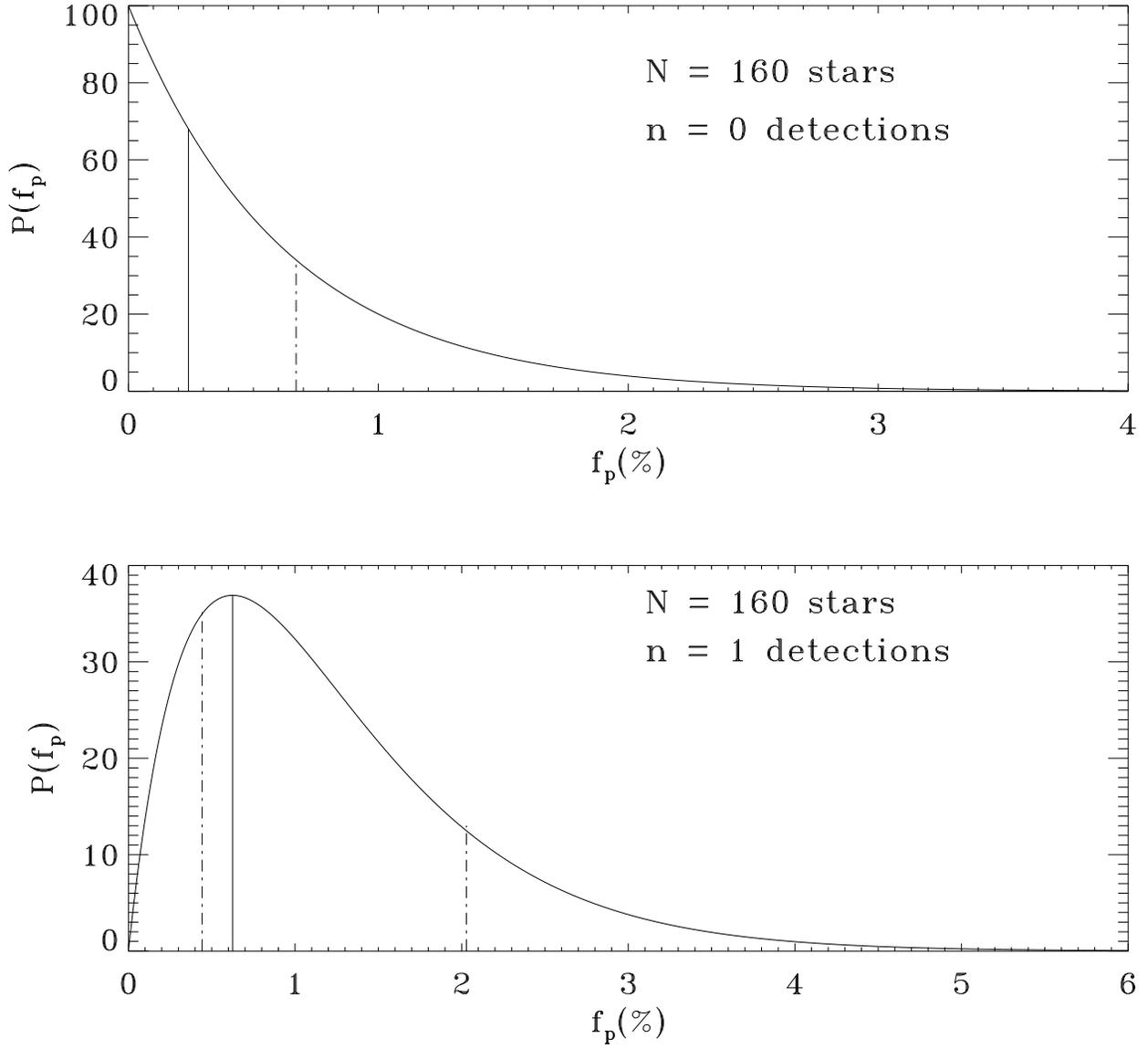}
\caption{Probability function for a given sample size and true companion frequency, 
and the case of no detections (upper panel) and one successful detection (lower panel). \label{binom}}
\end{figure}

The probability distribution is not symmetric about its maximum, thus we 
report the range in $f_p$ that delimits 68\% of the integrated probability 
function, equivalent to 1-$\sigma$ limits for a Gaussian distribution. We 
make two assumptions. First, based on the null result presented 
above, we set $n = 0$ and $N = 160$ (Figure~\ref{binom}, top panel), 
and find $f_p = 0.24^{+0.43}_{-0.24}\%$. Thus, at the 1-$\sigma$ confidence level, 
we constrain the frequency of giant planets orbiting metal-poor stars to be $f_p<0.67\%$. 
If we make the hypothesis that indeed we missed one planet with 
$K > 100$ m $s^{-1}$ and $P < 3$ yr, $e < 0.3$, then $n = 1$, and we 
find that there is a 68\% chance of $f_p$ ranging between 0.44\% and 2.0\%, 
thus $f_p = 0.62^{+1.38}_{-0.18}\%$. 

\section{Summary, discussion and conclusions}

We have presented an analysis of three years of precision Doppler measurements of 
160 metal-poor stars from our Keck/HIRES radial-velocity survey. 

We have performed several statistical tests for excess variability and long-term trends. 
Our sample contains nine clear RV variable stars (identified on the basis of the accordance 
between different statistical indicators). Six of them exhibit 
significant long-term linear trends, indicative of companions in the stellar/sub-stellar 
mass regime on very long periods ($P$ much larger than the survey duration $T$). 
After subtraction of these long-term trends, the RV residuals are consistent with our measurement errors. 
This sample is currently the subject of a direct imaging search in the IR 
to better characterize the actual nature of the detected companions, whose 
results will be reported in a forthcoming paper. As for the remaining three stars 
which appear to be RV variables, their non-linear 
RV variations are again compatible with the presence of long-period massive sub-stellar 
or stellar companions, or due to some type of mechanism intrinsic to the star. 
Given the limited amount of data collected (typically, only a handful of observations 
per star over a time-span of three years), we can say very little about companions 
inducing RV variations close to the average internal errors we obtain ($\simeq 10$ m s$^{-1}$). 
Overall, our results indicate that none of the stars with clearly detected RV variations 
appear to be orbited by massive giant planet companions with periods shorter than the survey 
duration. 

Based on our RV data and numerical simulations of planetary signatures, 
we have placed quantitative lower-mass limits for sub-stellar companions 
orbiting our metal-poor sample, as a function of orbital separation. 
Our survey would have detected, with a 99.5\% confidence level, virtually 
all companions on low-eccentricity orbits ($e < 0.3$) with velocity semi-amplitude 
$K \gtrsim 100$ m s$^{-1}$, or $M_c\sin i\gtrsim 3.0\,M_\mathrm{J}(P/\mathrm{yr})^{(1/3)}$, 
and orbital periods $P\lesssim 3$ yr. 
None of the stars in our sample exhibits radial-velocity variations compatible 
with the presence of companions inducing $K > 100$ m s$^{-1}$ and 
with periods shorter than the survey duration. The observed dearth of gas giant planets within 2 AU 
of metal-poor stars ($-2.0 <$[Fe/H]$< -0.6$), allows us to confirm and extend 
previous findings (e.g., Santos et al. 2004; Fischer \& Valenti 2005). 
The resulting average frequency of gas giants orbiting metal-poor dwarfs with 
$-2.0\lesssim$[Fe/H]$\lesssim -0.6$ is $f_p<0.67\%$ (at the $1-\sigma$ level). 
We note that our null result also provides supporting evidence for 
the existence of a `brown dwarf desert' (e.g., ) which extends to low metallicity 
stars. Indeed, current estimates from ongoing Doppler surveys imply a frequency $f_\mathrm{BD}=0.5\pm0.2\%$ 
of close-in ($a\leq 3$ AU) brown-dwarf companions to solar-type stars with [Fe/H]$\gtrsim 0.0$ 
(see for example Marcy et al. 2005). 
However, whether $f_\mathrm{BD}$ correlates with stellar metallicity, depending on the 
range of orbital separations considered, is presently a still poorly understood issue, 
and we will address this point in detail in a forthcoming paper which will describe the 
results of our IR imaging search for companions around metal-poor stars with RV trends. 

We are now in a position to put the null result from our Doppler survey 
in the context of the observed correlation between the rate of occurrence $f_p$ of giant 
planets and the metallicity of their main-sequence solar-type stellar hosts. 
The fact that metal-rich stars have a higher probability of harboring a giant 
planet than their low-metallicity counterparts has been quantified 
in particular by Santos et al. (2004) and Fischer \& Valenti (2005) (for a review 
of the issue, see for example Udry \& Santos 2007). Using all the available 
data at that time, with no selection criteria based on orbital period, RV amplitude, 
planet mass, and metallicity, Santos et al. (2004) found that $f_p\propto$ [Fe/H] for 
metallicities above solar, while $f_p\sim\,const.$ for metallicities below solar. 
Fischer \& Valenti (2005) constructed a sample of stars from their Doppler survey 
with uniform detectability threshold ($K > 30$ m s$^{-1}$, $P < 4$ yr), and found 
an even steeper dependence of $f_p$ on [Fe/H], fitting a power-law $f_p = 0.03\times10^{2.0[\mathrm{Fe/H}]}$ 
in the metallicity range $-0.5<$[Fe/H]$<0.5$. Both analyses concluded that 
at least 25\% of the stars with twice the metal content of our Sun ([Fe/H] = 0.3) 
are orbited by a giant planet ($M_p\gtrsim 1$ $M_J$), and this number decreases to 
below 5\% for solar-neighborhood dwarfs with [Fe/H]$\lesssim 0.0$. 

\begin{figure}
\plotone{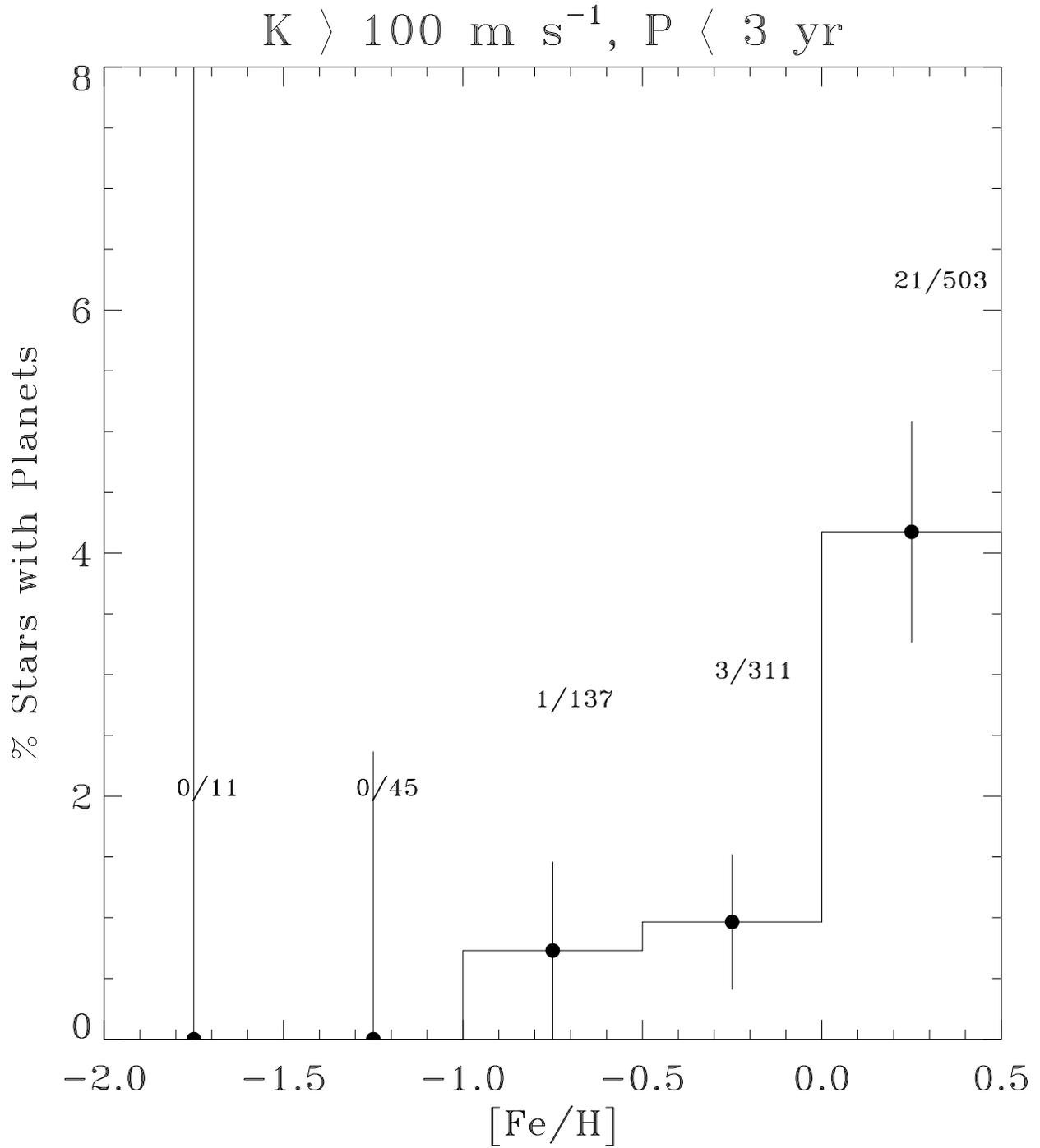}
\caption{Percentage of giant-planet hosting stars as a function of metallicity (0.5 dex bins) for 
the sample constructed combining our survey stars with those from the Fischer \& Valenti (2005) 
database that have $K > 100$ m s$^{-1}$ and $P < 3$ yr.
\label{dist_feh1}}
\end{figure}

We have combined our dataset with the uniform sample of Fischer \& Valenti (2005), 
taking care to select objects from the latter sample with planets that would have 
been detected by both surveys, i.e. with $K > 100$ m s$^{-1}$ and $P < 3$ yr. In 
Figure~\ref{dist_feh1} we present the fraction of planet-hosting stars as function 
of [Fe/H] for the combined sample. The ratio of planet-bearing stars to all stars 
is labeled above each 0.5-dex bin in metallicity. Error bars for [Fe/H]$> -1.0$  
are calculated assuming Poisson statistics (i.e., the percentage of stars with planets 
divided by the square root of the number of planets), while, given the small-number 
statistics, a binomial distribution is utilized to determined the $1-\sigma$ error bars 
for [Fe/H]$< -1.0$. In the bin $-1.0<$[Fe/H]$<-0.5$ the merging of the two datasets 
provides for a five-fold increase in the number of stars surveyed with respect to the 
analogous analysis reported in Figure 4 of Fischer \& Valenti (2005). Overall, 
the probability of forming gas giant planets with $K > 100$ m s$^{-1}$ and $P < 3$ yr 
around metal-poor stars ([Fe/H]$< 0.0$) appears suppressed by a factor of several with respect to 
that of finding any around metal-rich stars ([Fe/H]$> 0.0$). 

\begin{figure}
\plotone{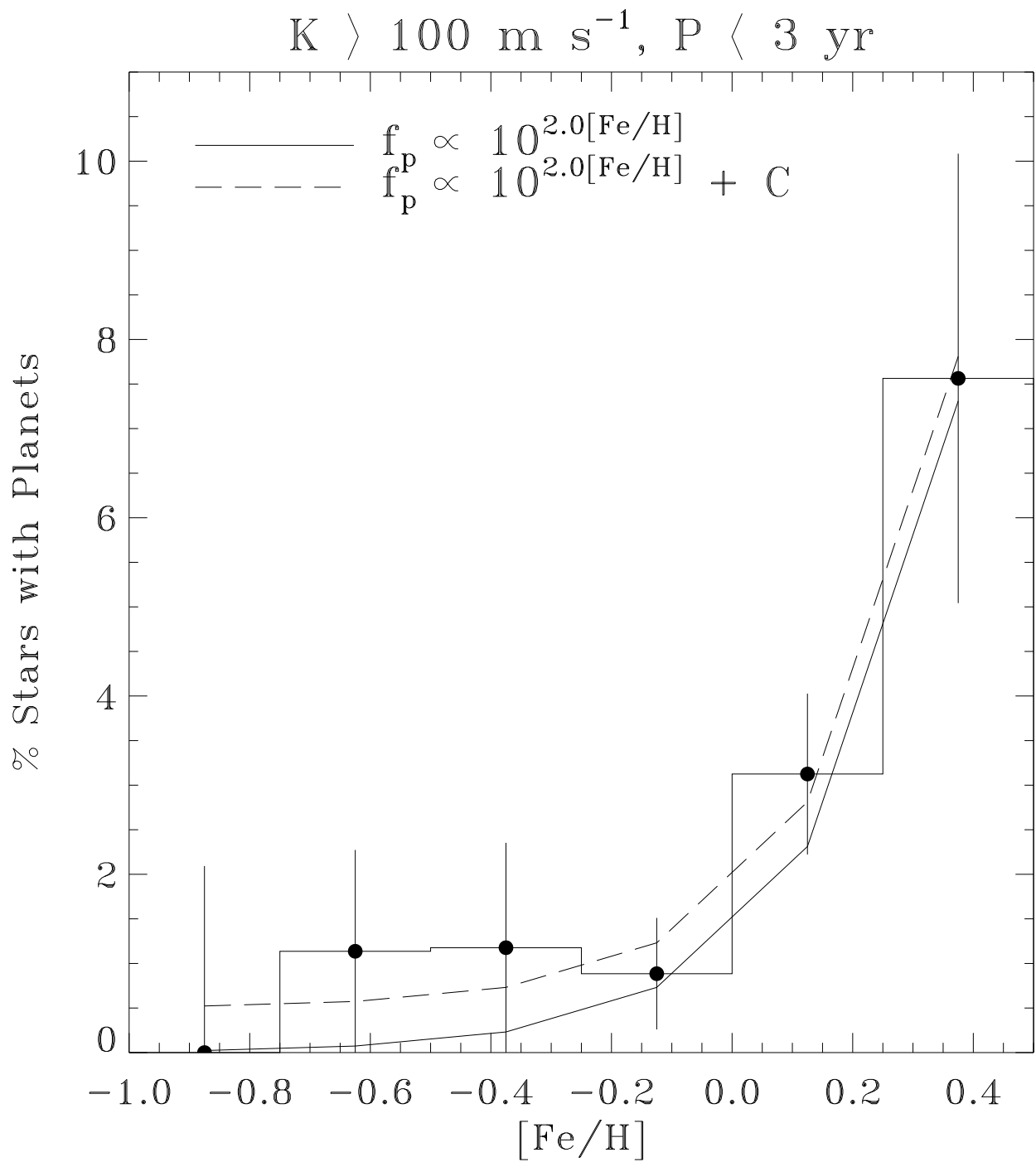}
\caption{Same as Figure~\ref{dist_feh1}, but divided into 0.25 dex metallicity bins. The
increasing trend in the fraction of stars with planets as a function of 
metallicity is well fitted with a power law, but the data are compatible with a
constant occurrence rate $f_p\simeq 1\%$ for [Fe/H]$\lesssim 0.0$.\label{dist_feh2}}
\end{figure}

To investigate this correlation in more detail, we divide the stars into finer 
0.25-dex metallicity bins, and limit ourselves to the metallicity range $-1.0<$[Fe/H]$<0.5$, 
for which better statistics are available. Figure~\ref{dist_feh2} shows 
a smooth increase in the fraction of stars with planets
as a function of increasing metallicity above [Fe/H]$ = 0.0$, and an approximately 
constant detection rate for metallicity below solar. About 7.5\% of stars with [Fe/H]$\simeq 0.3$ 
host planets with $K > 100$ m s$^{-1}$ and $P < 3$ yr, but $f_p\sim 1.0\%$ for companions 
with the same characteristics orbiting sub-solar metallicity stars. The two curves 
quantify the $f_p-$[Fe/H] relationship in terms of a power-law with the same 
index as the Fischer \& Valenti (2005) functional form, with and without a constant term: 

\begin{equation}
f_p(\%) = 1.3\times 10^{2.0[\mathrm{Fe/H}]} (+0.5)
\end{equation}

We confirm that the likelihood of a star to harbor a planet more massive 
than Jupiter within 2 AU is a steeply rising function of the host's metal content. 
However, while $f_p(-1.0<$[Fe/H]$<-0.5)$ appears to be a factor of several lower than 
$f_p($[Fe/H]$>0.0)$, it is indistinguishable from $f_p(-0.5<$[Fe/H]$<0.0)$. 
Overall, the data for stars with $-1.0\lesssim$[Fe/H]$\lesssim 0.0$ are compatible, 
in a statistical sense, with a constant occurrence rate $f_p\simeq 1\%$, as 
the two power-law fits are statistically indistinguishable one from the other. The difficulty 
in representing the $f_p-$[Fe/H] relationship as a single power-law for the entire range 
of metallicities encompassed in this analysis might either indicate bi-modality in the 
distribution (flat behavior for [Fe/H]$< 0.0$ plus strong rise for [Fe/H]$> 0.0$), or simply 
constitute an as of yet not well-characterized low-metallicity tail. Similar conclusions 
are derived by Udry \& Santos (2007), who analyzed the average $f_p-$[Fe/H] in the 
California-Carnegie and Geneva samples. 
Our results can usefully inform theoretical studies of the process of giant planet formation 
across two orders of magnitude in metallicity. We summarize below the existing predictions, 
and then discuss how our data can contribute to test them. 

Core accretion models for giant planet formation are today mature enough to make clear 
predictions concerning the mass-range, orbital period distribution, and frequency of planets 
orbiting stars of varying metallicity (assuming their metal content closely track the 
metallicity of the protoplanetary disk in which they formed). In the standard model of 
gas giant planet formation by core accretion (e.g., Pollack et al. 1996; Alibert et al. 2005), 
several studies have highlighted how $f_p$ (integrated over a wide range of planetary masses) 
should be a monotonically decreasing function of decreasing [Fe/H] (e.g., Ida \& Lin 2004; 
Kornet et al. 2005; Matsuo et al. 2007). Indeed, Matsuo et al. (2007) propose that there should be an 
effective low-metallicity limit at [Fe/H]$\approx -1.0$ below which giant planets cannot 
form by core accretion at all. Those authors also suggest the existence of a metallicity dependent 
upper mass limit, of about 2 $M_J$ at [Fe/H]$\approx -1.0$. On the other hand, the likelihood 
of finding Neptune-mass planets might not correlate strongly with [Fe/H] of the host 
(Ida \& Lin 2004; Benz et al. 2006). Finally, short-period planets may not be found 
around metal-poor stars either because of an inverse metallicity-dependence of migration rates 
(Livio \& Pringle 2003) or due to longer timescales for giant planet formation around 
metal-poor stars, and thus reduced migration efficiency before the disk dissipates 
(Ida \& Lin 2004). As for the alternative scenario of giant planet formation by disk instability, 
the situation is less clear, even for the metallicity correlation (Boss 2002; Matsuo et
al. 2007; cf. Mayer et al. 2007). Estimated masses for gas giant planets formed by disk
instability range from Saturn-mass (Mayer et al. 2004; Boss 2008) to $\sim$ 10 Jupiter
masses (Boss 1998; Mayer et al. 2004). Other authors (Rice et al. 2003; 
Rafikov 2005) have concluded that giant planets formed by disk instability should 
populate preferably the high-mass tail of the mass distribution. Low migration efficiency 
in gravitationally unstable disks implies that planets formed by this mechanism 
should be found on wide orbits (Rice et al. 2003; Mayer et al. 2004; Rafikov 2005). 

We do not detect any massive planets ($M_p\sin i\gtrsim 1-4$ $M_J$) within 2 AU of metal-poor 
stars with $-2.0\lesssim$[Fe/H]$\lesssim -0.6$, and constrain their occurrence rate 
to be no larger than $f_p\simeq 1\%$. All long-period trends identified in our survey 
are compatible with being induced by brown dwarf or stellar companions. Based on the 
above expectations, these findings appear to be circumstantial evidence in favor of the core 
accretion mechanism of giant planet formation. On the other hand, the probability of a 
star with $-1.0\lesssim$[Fe/H]$\lesssim 0.0$ to host a massive planet within 2 AU 
appears constant, and this could be read as supporting evidence for the alternative 
disk instability model. However, though very useful, our results are not resolutive, and 
at least three observational avenues can be identified to expand and improve the statistics 
and thus further constrain proposed models. 

First, it is necessary to improve on the mass sensitivity threshold of the surveys. 
The formation of lower-mass giant planets around relatively low-metallicity 
stars is not completely inhibited, as recently testified by the announcements 
of a giant planet ($M_p\sin i = 1.8$ $M_J$) orbiting HD 171028 ([Fe/H]$= -0.49$, Santos et al. 2007) 
at an orbital distance of $\sim 1.3$ AU ($K\simeq 60$ m s$^{-1}$) and of a two-planet system 
($M_p\sin i\simeq 0.9$ $M_J$ and $M_p\sin \simeq 0.5$ $M_J$) around 
HD 155358 ([Fe/H]$ = -0.65$, Cochran et al. 2007), with orbital radii between 
$\sim0.6$ AU and $\sim 1.3$ AU ($K\simeq 40$ m s$^{-1}$ and $K\simeq 15$ m s$^{-1}$, 
respectively). None of these stars were included in our Keck survey either because of the [Fe/H] 
cut-off, or due to other selection criteria (proper motion, declination limits) in 
the original Carney et al. (1994) and Ryan (1989) surveys from which our sample 
was drawn. 
However, the evidence for a few objects in our sample with apparently short-term rms RV variations 
exceeding 2 to 3 times the measurements uncertainties indicates that at lower 
masses the fraction of metal-poor stars hosting planets may be larger. 
Unfortunately, with our data we cannot confirm or rule out the early evidence for a vanishing $f_p-$[Fe/H] 
correlation in the Saturn to Neptune companion mass regime, argued for by 
Udry \& Santos (2007) and more recently corroborated by the findings of Sousa et al. (2008). 

Second, studies of planet occurrence around metal-poor stars still suffer from 
small-number statistics biases, and one would clearly gain further insight by 
expanding the sample size. Overall, only $200-300$ objects are presently targeted 
from the ground in the metallicity regime $-2.0\lesssim$[Fe/H]$\lesssim-0.5$, and when statistical analyses 
such as ours are carried out, usually only several tens of stars end up populating a 
given metallicity bin. In particular, at the present time we still cannot completely 
rule out the possibility that the evidence for bimodality in the metallicity distribution 
is simply due to the limited number of stars surveyed. 

Finally, it would be very beneficial to extend the time baseline of the 
surveys, to probe the existence of longer period planets. This is the only way 
to assess the relative role of metallicity in migration processes, 
as different conclusions would be drawn in the event massive planets 
are found or not at orbital radii of several AUs. 

While challenging, the above experiments may be successfully carried out by 
ongoing RV programs (e.g., Santos et al. 2007), next-generation RV surveys (e.g., Ge et al. 2007), 
upcoming ultra-high precision transit surveys in space such as Kepler (Borucki et al. 2003), 
and future high-precision space-borne astrometric observatories such as Gaia (Casertano et al. 2008), 
SIM/SIM-Lite (Unwin et al. 2008). These data are crucially needed in order to better our 
understanding of the complex interplay between planet formation processes, migration scenarios, 
and several important characteristics of the protoplanetary disks, such as their metal content.

\acknowledgments

An anonymous referee provided useful and timely comments and suggestions which helped 
improve the paper. GT acknowledges partial support for
this work from NASA Origins grant NNG04LG89G. AS gratefully acknowledges the
Kepler mission for partial support under NASA Cooperative Agreement
NCC 2-1390. JBL was supported for this work by NSF grant AST-0307340. 
The data presented herein were obtained at the
W.M. Keck Observatory, which is operated as a scientific partnership
among the California Institute of Technology, the University of
California and the National Aeronautics and Space Administration. The
Observatory was made possible by the generous financial support of the
W.M. Keck Foundation. The authors wish to recognize and acknowledge
the very significant cultural role and reverence that the summit of
Mauna Kea has always had within the indigenous Hawaiian community. We
are most fortunate to have the opportunity to conduct observations
from this mountain. This research has made use of NASA's Astrophysics
Data System Abstract Service and of the SIMBAD database, operated at
CDS, Strasbourg, France.



\clearpage

\begin{deluxetable}{llllccccccccc}
\tabletypesize{\tiny}
\tablecaption{Target list of the Keck HIRES metal-poor planet survey\label{taball}}
\tablewidth{0pt}
\tablehead{\colhead{Star Name} & \colhead{Alt. Name}&\colhead{RA} & \colhead{DEC} & \colhead{$V$} &
           \colhead{$T_\mathrm{eff}$} & \colhead{[Fe/H]}&\colhead{$M_\star$}&\colhead{$B-V$}&\colhead{$V-K$}&
           \colhead{$\log g$} & \colhead{$d$ (phot)} & \colhead{$d$ (Hip)}\\
           \colhead{} & \colhead{(HD/HIP)}&\colhead{(J2000)} &  \colhead{(J2000)} & \colhead{(mag)} & \colhead{(K)} & 
           \colhead{(dex)}& \colhead{$(M_\odot)$}&
           \colhead{} & \colhead{} & \colhead{(dex)} & \colhead{(pc)} & \colhead{(pc)}}
\startdata
     G 30-56\dotfill & HIP 1301\dotfill  &  0 16 20.87 & +16 38 46.9&  9.73  & 5753  & $-$0.66 &  0.89 &  0.57 &  1.49 &  4.47 &    103 &    115 \\
   G 158-102\dotfill &  &  0 35 41.26 & $-$06 42 36.6& 10.82  & 5320  & $-$0.73 &  0.78 &  0.70 &  1.75 &  4.60 &    113 &  \dots \\
     HD 3567\dotfill &  &  0 38 31.95 & $-$08 18 33.4&  9.25  & 6010  & $-$1.34 &  0.83 &  0.46 &  1.35 &  4.36 &     98 &  104 \\
    G 270-49\dotfill & HIP 3441\dotfill  &  0 43 57.09 & $-$07 32 01.0&  9.38  & 5658  & $-$0.66 &  0.85 &  0.58 &  1.56 &  4.53 &     84 &   91 \\
     G 70-17\dotfill &  &  0 50  6.50 & $-$04 25 55.8& 11.29  & 4828  & $-$0.82 &  0.64 &  0.88 &  2.30 &  4.71 &     84 &   \dots \\
      G 2-19\dotfill & HD 6500\dotfill  &  1  6  5.15 & +01 42 23.1&  9.46  & 5618  & $-$0.60 &  0.86 &  0.60 &  1.57 &  4.53 &     83 &     79 \\
     HD 7424\dotfill &  &  1 14  7.24 & $-$16 25 35.0& 10.08  & 5538  & $-$0.76 &  0.82 &  0.67 &  1.64 &  4.56 &     97 &    116 \\
    G 271-34\dotfill & HD 7983\dotfill  &  1 19  0.00 & $-$08 56 22.2&  8.90  & 5638  & $-$0.73 &  0.85 &  0.59 &  1.53 &  4.52 &     65 &     67 \\
\enddata




\tablecomments{Target list of the Keck/HIRES metal-poor planet search program. We provide
visual magnitude, effective temperature, metallicity, mass, colors, surface gravity, and 
distance estimates}
\end{deluxetable} 

\clearpage

\begin{deluxetable}{lcccc}
\tablecaption{Radial-velocity results for all stars}
\tablewidth{0pt} 
\tablehead{\colhead{Star Name} & 
           \colhead{$N$} & \colhead{$\sigma_\mathrm{rms}$} & \colhead{$\bar\sigma_\mathrm{int}$} & 
           \colhead{$\Delta T$}\\
           \colhead{} & \colhead{} & 
           \colhead{(m s$^{-1}$)} & \colhead{(m s$^{-1}$)} & \colhead{(days)}}
\startdata
     G30-56\dotfill &  5  &  21.4  &   6.9 &   378.0\\
   G158-102\dotfill &  5  &   7.2  &  10.2 &   378.0\\
     HD3567\dotfill &  3  &  11.1  &   8.8 &   115.7\\
    G270-49\dotfill &  5  &  10.5  &   9.8 &   378.0\\
     G70-17\dotfill &  5  &   7.7  &  11.7 &   378.0\\
      G2-19\dotfill &  5  &  12.9  &  11.7 &   377.9\\
     HD7424\dotfill &  6  & 236.0  &   6.7 &   492.7\\
    G271-34\dotfill &  5  &  13.6  &  11.1 &   377.0\\
\enddata




\tablecomments{Radial-velocity results for all stars observed at least twice 
during the survey. For each star, we list the total number of spectra collected $N$, 
the total RV scatter $\sigma_\mathrm{rms}$, the average internal measurement error 
$\bar\sigma_\mathrm{int}$, and the time-span of the observations $\Delta T$\label{totrv}}
\end{deluxetable}

\clearpage

\begin{deluxetable}{lccccc}
\tablecaption{Variability analysis for the sample}
\tablewidth{0pt} 
\tablehead{\colhead{Star Name} & 
           \colhead{$N$} & 
           \colhead{$\left(\frac{\displaystyle\sigma_\mathrm{rms}}{\displaystyle\bar\sigma_\mathrm{int}}\right)_\mathrm{const}$} & 
           \colhead{Pr($\chi^2_\nu$)} & 
           \colhead{Pr($Ku$)}&\colhead{Pr($F$)}}
\startdata
         G30-56\dotfill &     5  &    3.083  &      0.00000  &      0.03179  &  6.17e-05\\
       G158-102\dotfill &     5  &    0.713  &      0.69393  &       0.60753  &      0.97951\\
         HD3567\dotfill &     3  &    1.251  &     0.070947  &        \dots   &      0.28643\\
        G270-49\dotfill &     5  &    1.073  &      0.28123  &       0.28040  &      0.35165\\
         G70-17\dotfill &     5  &    0.657  &      0.64313  &       0.91370  &      0.77199\\
          G2-19\dotfill &     5  &    1.102  &     0.04342  &       0.33624  &     0.03304\\
         HD7424\dotfill &     6  &   35.169  &      0.00000  &    0.00014  &  2.80e-45\\
        G271-34\dotfill &     5  &    1.224  &      0.10093  &      0.07288  &     0.06761\\
\enddata




\tablecomments{$\chi^2$-test, $F$-test, and $Ku$-test for the complete sample. 
\label{stattests} }
\end{deluxetable}

\newpage
      
\begin{deluxetable}{lccccccc}
\tablecaption{Test for long-term trends}
\tablewidth{0pt} 
\tablehead{\colhead{Star Name} & 
           \colhead{$N$} & \colhead{$\left(\frac{\displaystyle\sigma_\mathrm{rms}}{\displaystyle\bar\sigma_\mathrm{int}}\right)_\mathrm{const}$} 
           &\multicolumn{1}{c}{Slope}& \colhead{$F$}&\colhead{P($F$)}&
           \colhead{$\left(\frac{\displaystyle\sigma_\mathrm{rms}}{\displaystyle\bar\sigma_\mathrm{int}}\right)_\mathrm{lin}$}&
           \colhead{$Pr(\chi^2_\nu)_\mathrm{lin}$} \\
           \colhead{}&\colhead{}&\colhead{}&\multicolumn{1}{c}{m s$^{-1}$ yr$^{-1}$}&\colhead{}&
           \colhead{}&\colhead{}&\colhead{}}
\startdata
      G30-56\dotfill &   5 &   3.08&   $-$35.63$\pm$5.87 &       4.8 &    0.15802 &         1.57&   0.00815\\
    G158-102\dotfill &   5 &   0.71&      4.23$\pm$8.83 &       1.3 &    0.82039 &        0.71&     0.61624\\
      HD3567\dotfill &   3 &   1.25&     74.43$\pm$ 35.57 &       7.3 &    0.24196 &        0.57&     0.41505\\
     G270-49\dotfill &   5 &   1.07&   $-$19.16$\pm$  9.23 &       9.3 &   0.05290 &        0.39&     0.87903\\
      G70-17\dotfill &   5 &   0.66&      2.02$\pm$ 10.11 &       1.0 &    0.99317 &        0.73&     0.50310\\
       G2-19\dotfill &   5 &   1.10&   $-$24.25$\pm$ 10.65 &       2.2 &    0.45587 &        0.82&     0.30685\\
      HD7424\dotfill &   6 &  35.17&  $-$417.02$\pm$  4.66 &     1000. & 3.28e-07 &         1.21&     0.17720\\
     G271-34\dotfill &   5 &   1.22&   $-$25.12$\pm$  9.86 &       13. &   0.02939 &        0.38&     0.81465\\
\enddata




\tablecomments{Linear RV trend tests for our sample. For each star with more than 
three observations, we report the values of $(\sqrt{\chi^2_\nu})_\mathrm{const}$ 
(for comparison from Table~\ref{stattests}), 
the slope and its formal error of a linear fit to the RV data, the $F$-test results, and 
the recomputed $(\sqrt{\chi^2_\nu})_\mathrm{lin}$ and $Pr(\chi^2_\nu)_\mathrm{lin}$ 
on the post-fit residuals.  \label{trendstest} }
\end{deluxetable}

\newpage
\newpage
\begin{deluxetable}{lrr}
\tablecaption{Radial velocity measurements for stars with $\sigma_\mathrm{rms}/\bar\sigma_\mathrm{int}\lesssim 3.0$.
\label{rvsumm}}
\tablewidth{0pt} 
\tablehead{\colhead{BJD $-$ 2,400,000} & \colhead{Radial Velocity} & \colhead{$\sigma_\mathrm{RV}$} \\
\colhead{}& \colhead{(m s$^{-1}$)} & \colhead{(m s$^{-1}$)}}
\startdata
\noalign{\vskip 0.2cm}
\multicolumn{3}{c}{G 30-56} \\
\noalign{\vskip 0.3cm}
53192.05224  &   0.4  & 5.5 \\
53209.09366  &  28.3  & 7.0 \\
53210.07438  &  18.7  & 8.8 \\
53569.08494  & $-$24.4  & 5.3 \\
53570.07546  & $-$23.1  & 7.9 \\
\noalign{\vskip 0.2cm}
\multicolumn{3}{c}{G 158-102} \\
\noalign{\vskip 0.3cm}
53192.07375  & $-$14.3  & 12.2 \\
53209.07767  &   1.6  & 12.1 \\
53210.05681  &   4.0  & 6.0 \\
53569.10203  &   5.1  & 9.6 \\
53570.09542  &   3.6  & 10.8 \\
 \enddata
\tablecomments{We intend for this table to appear in its 
entirety in the electronic version of the journal. A portion is 
shown here to illustrate its format. The data are also available from the authors upon request.}
\end{deluxetable}
\newpage

\begin{deluxetable}{lrr}
\tablecaption{Radial velocity measurements for stars with linear trends.\label{rvlin}}
\tablewidth{0pt} 
\tablehead{\colhead{BJD - 2,400,000} & \colhead{Radial Velocity} & \colhead{$\sigma_\mathrm{RV}$} \\
\colhead{}& \colhead{(m s$^{-1}$)} & \colhead{(m s$^{-1}$)}}
\startdata
\noalign{\vskip 0.2cm}
\multicolumn{3}{c}{HD 7424} \\
\noalign{\vskip 0.3cm}
53192.10039  &  241.1 &   6.4\\
53209.12624  &  225.2 &   3.7\\
53210.11312  &  224.9 &   7.6\\
53569.11138  & $-180.7$ &   8.5\\
53570.10677  & $-178.1$ &   6.0\\
53684.76920  & $-332.4$ &   7.8\\

\noalign{\vskip 0.2cm}
\multicolumn{3}{c}{G 237-84} \\
\enddata
\end{deluxetable}


\begin{deluxetable}{lrr}
\tablecaption{Radial velocity measurements for stars with non-linear trends.\label{rvnolin}}
\tablewidth{0pt} 
\tablehead{\colhead{BJD - 2,400,000} & \colhead{Radial Velocity} & \colhead{$\sigma_\mathrm{RV}$} \\
\colhead{}& \colhead{(m s$^{-1}$)} & \colhead{(m s$^{-1}$)}}
\startdata
\noalign{\vskip 0.3cm}
\multicolumn{3}{c}{HD 210295} \\
\noalign{\vskip 0.3cm}

53209.04688  &    9.4 &   5.5\\
53210.03125  &   64.6 &  10.7\\
53569.00391  &   37.0 &   8.7\\
53569.98828  &   18.8 &  15.8\\
53949.98438  &  $-$50.0 &   7.1\\
53951.05469  &  $-$59.6 &   8.8\\
53952.08203  &  $-$20.2 &   8.4\\
\noalign{\vskip 0.3cm}
\multicolumn{3}{c}{G 27-44} \\
\noalign{\vskip 0.3cm}

50668.01562  &   31.5 &   7.3\\
50670.01562  &   32.2 &   8.9\\
51008.06641  &   41.4 &   3.7\\
51038.98438  &   22.9 &   7.5\\
51545.72266  &   26.6 &  14.5\\
53162.06250  &    0.0 &   6.4\\
53569.00781  &  $-$80.1 &   8.4\\
53952.08984  &  $-$74.7 &  10.7\\
\noalign{\vskip 0.3cm}
\multicolumn{3}{c}{G 28-43} \\
\noalign{\vskip 0.3cm}

53162.08430 &  $-$401.4 &   7.3\\
53209.05709 &    51.8 &  43.9\\
53210.03967 &   164.1 &  38.5\\
53569.03491 &   402.8 &  18.5\\
53570.02176 &   449.4 &  25.5\\
53684.75181 &  $-$666.8 &  46.1\\

 \enddata
\end{deluxetable}

\newpage

\newpage

\newpage

\newpage

\end{document}